\documentclass{PoS}
\usepackage{natbib}
\usepackage{journals}

\title{Star-formation in the host galaxies of radio-AGN}

\ShortTitle{Star-formation in the host galaxies of radio-AGN}

\author{\speaker{Marios Karouzos}\\

        CEOU, Seoul National University, Seoul, Republic of Korea\\

        E-mail: \email{mkarouzos@astro.snu.ac.kr}}

\author{Markos Trichas\\

	Harvard-Smithsonian Center for Astrophysics, Cambridge, MA 02138, USA}

\author{Myungshin Im\\

	CEOU, Seoul National University, Seoul, Republic of Korea}

\author{Matt Malkan\\

	Department of Physics \& Astronomy, UCLA, Los Angeles, CA 90095-1547, USA}

\author{and the AKARI-NEP team}

\abstract{There exist strong evidence supporting the co-evolution of central supermassive black holes and their host galaxies. It is however still unclear what the exact role of nuclear activity, in the form of accretion onto these supermassive black holes, in this co-evolution is. We use a rich multi-wavelength dataset available for the North Ecliptic Pole field, most notably surveyed by the AKARI satellite infrared telescope to study the host galaxy properties of AGN. In particular we are interested in investigating star-formation in the host galaxies of radio-AGN and the putative radio feedback mechanism, potentially responsible for the eventual quenching of star-formation. Using both broadband SED modeling and optical spectroscopy, we simultaneously study the nuclear and host galaxy components of our sources, as a function of their radio luminosity, bolometric luminosity, and radio-loudness. Here we present preliminary results concerning the AGN content of the radio sources in this field, while offering tentative evidence that jets are inefficient star-formation quenchers, except in their most powerful state.}

\FullConference{Nuclei of Seyfert galaxies and QSOs - Central engine \& conditions of star formation\\

                 November 6-8, 2012\\

                 Max-Planck-Insitut f\"ur Radioastronomie (MPIfR), Bonn, Germany}

\begin{document}

\section{Introduction}
\label{sec:intro}
The discovery of a number of scaling relations (e.g., \citealt{Ferrarese2000}, \citealt{Magorrian1998}, \citealt{Kormendy1995}) connecting nuclear properties of galaxies to their global characteristics has led to an ongoing debate about the possible physical processes that give rise to this connection.\\
One such candidate physical process is feedback. The presence of activity within the nucleus of the galaxy (AGN) can give rise to outflows that may affect the whole galaxy. In particular signatures of star-formation quenching should be found in the host galaxies of radio-AGN, as radio-loud active galaxies are characterized by powerful, well-collimated outflows.\\
AGN radio-jets are known to be able to deposit large quantities of mechanical energy in their surroundings (e.g., \citealt{McNamara2005}). The role of radio-loud AGN and their jets with respect to star-formation has been studied intensively (e.g., \citealt{Bicknell2000}, \citealt{Croton2006}, \citealp{Best2007,Best2012}, \citealt{Kalfountzou2012}). Results have been ambiguous, with evidence for both negative and positive impact of radio-AGN outflows on star-formation.\\
We therefore investigate the broadband spectral energy distributions (SEDs) of a sample of radio sources and try to decouple, by means of SED template fitting, the AGN and star-formation components. We use the cosmological parameters $H_{0}=71$ $km s^{-1} Mpc ^{-1}$, $\Omega_{M}=0.27$, and $\Omega_{\Lambda}=0.73$ (from the first-year WMAP observations; \citealt{Spergel2003}).

\section{The AKARI North Ecliptic Pole (NEP) field}
\label{sec:NEP}
The AKARI space telescope (\citealt{Murakami2007}), using the InfraRed Camera (\citealt{Onaka2007}) carried out observations of legacy fields, achieving greater sensitivity and resolution than previous instruments. One of these is the North Ecliptic Pole (NEP) field, which totals an area of $\sim5.4$ deg$^{2}$ (\citealt{Kim2012}). With 9 spectral bands, ranging from 2.4$\mu$m to 24$\mu$m, the IRC continuously covers the whole near- to mid-IR wavelength range, including the prominent wavelength gap (9-20$\mu$m) that characterized Spitzer observations.\\
For the NEP-Wide field exist also: deep GALEX observations (PI: Malkan), deep optical observations (CFHT; \citealt{Hwang2007}, Maidanak; \citealt{Jeon2010}), near-IR observations (FLAMINGOS; Jeon et al., in prep.), and radio observations at 1.4 GHz (WSRT;  \citealt{White2010}). The all-sky survey of the WISE telescope (\citealt{Wright2010}) can provide us with additional data in the near- to mid-IR regime. Finally, several spectroscopic campaigns have also taken place (with WYIN, MMT, and Keck telescopes). In total more than 2000 spectroscopic redshifts are available in the NEP-Wide field (Shim et al., in prep.).\\

\section{Sample and method}
\label{sec:sample}
\subsection{Radio-IR source cross-matching}
We cross-match the AKARI catalog of \citet{Kim2012} with the 1.5 GHz catalog of \citet{White2010}. The original IR band-merged AKARI NEP-Wide catalog contains 114794 sources at an N2 (2.4$\mu$m) band AB magnitude limit of $\sim21$ and a resolution of $\sim4$ arcseconds. For the cross-matching we use a sub-sample of 107504 near-IR detected sources (5$\sigma$ detection in N2 or N3 bands).  The radio catalog covers an area of 1.7 deg$^{2}$ with a beam size of 17 arcseconds and a sensitivity of 21 $\mu$ Jy/beam. The final catalog contains 462 radio sources at a 5$\sigma$ detection limit.\\
To cross-match the two catalogs, we use the Poisson-probability based method of \citet{Downes1986} that takes into account both the proximity and apparent magnitude of an infrared neighbor source in order to calculate the probability of a true match. The final IR-radio catalog contains 425 cross-matched sources. 

\subsection{Photometric redshifts and radio-loudness}
\begin{figure}[htbp]
\begin{center}
\includegraphics[width=0.49\textwidth,angle=0]{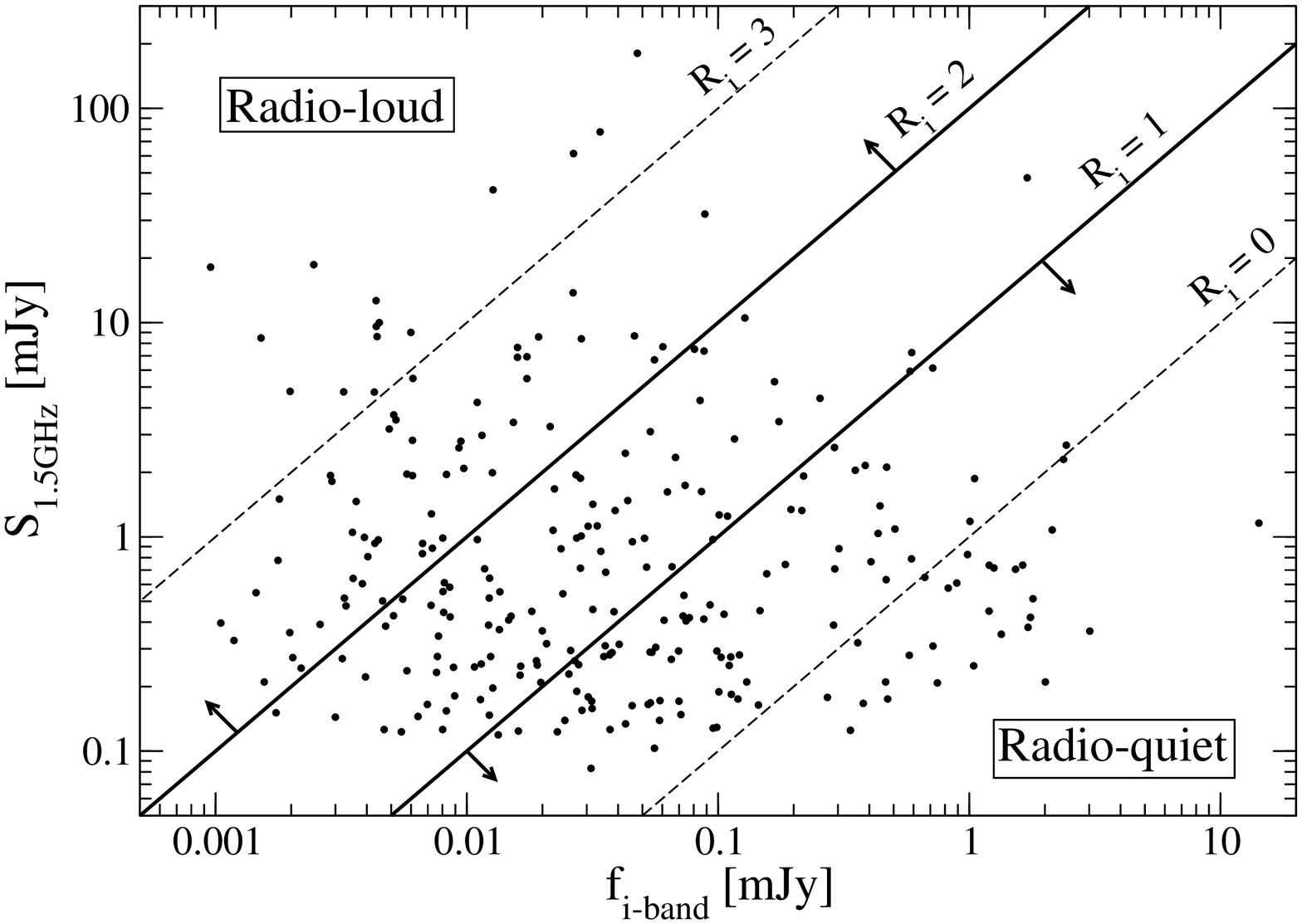}
\includegraphics[width=0.425\textwidth,angle=0]{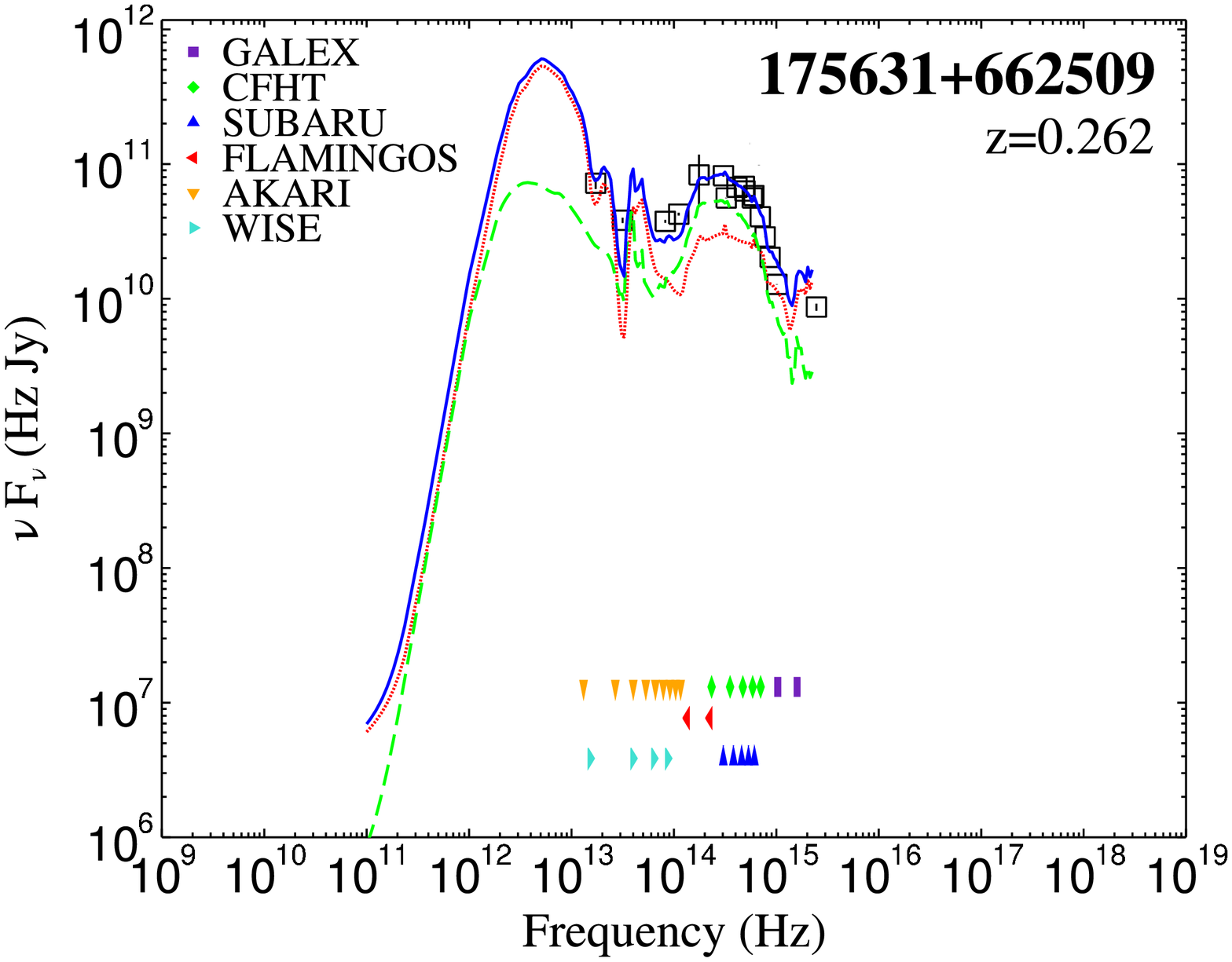}
\caption{Left: Extinction-corrected optical i-band flux and radio 1.5 GHz flux for the sample of 266 IR-radio sources with acceptable photometric redshifts. The diagonal lines denote constant radio-loudness values, with the solid lines marking the limits for radio-loud and radio-quiet sources. Right: Example of a source SED and the total fit (blue) with both an AGN (green) and star-formation (red) contribution.}
\label{fig:opt_radio}
\end{center}
\end{figure}

We calculate photometric redshifts for the IR-radio cross-matched sources using the publicly available LePhare code (\citealt{Arnouts1999}, \citealt{Ilbert2006}). We use the near-UV GALEX band as well as the full optical bands and near-IR bands, extending out to the W2 WISE band (4.6 $\mu$m). For the photometric redshift estimation we use the set of CFHT galactic SED templates from \citet{Ilbert2006}, the \citet{Polletta2007} AGN templates, and several stellar template libraries. After excluding stars and sources for which LePhare failed to determine a photometric redshift, we are left with 266 sources with acceptable photometric redshifts. Given the limitations of photometric redshifts for the following we consider sources at $z\leqslant2.0$.\\
\\
Originally radio-loudness, $R_{i}$, is defined as the ratio of the luminosities at 5 GHz and 4000\AA. Under that definition, radio-loud sources are those with $R_{i}>10$. Given the available data for the NEP-Wide field, here we use an alternative definition of radio-loudness from \citet{Ivezic2002}, utilizing the luminosities at 1.5 GHz and the CFHT i-band (7000\AA). Under that definition, we classify sources with $R_{i}>2$ as radio-loud, while for $R_{i}<1$ a source is classified as radio-quiet. For the calculation of the radio-loudness we are using i-band fluxes corrected for Galactic extinction (from \citealt{Schlegel1998}). In Fig. \ref{fig:opt_radio} (left) we show the distribution of optical i-band and radio 1.5 GHz fluxes, together with constant radio-loudness lines. According to our definitions above, we have in total 70 radio-loud and 105 radio-quiet sources.

\subsection{SED fitting}
Using the full broadband SEDs (ranging from UV to mid-IR) we employ the method and templates of \citet{Trichas2012} to fit sources simultaneously and additively with both an AGN and star-forming template. In this way, for each source the total bolometric luminosity, $L_{bol}$, and the fractional bolometric contribution $\alpha$ of the AGN component are derived. An example of such a fit is shown in Fig. \ref{fig:opt_radio} (right). Through $\chi^{2}$ and eye evaluation of the fits we define a sub-sample of 156 for which we have satisfactory a SED fit, of which 145 have redshifts $z\leqslant2.0$. We use these for the following.\\

\section{Results}
\subsection{AGN contribition vs. $L_{1.5GHz}$}
\begin{figure}[htbp]
\begin{center}
\includegraphics[width=0.49\textwidth,angle=0]{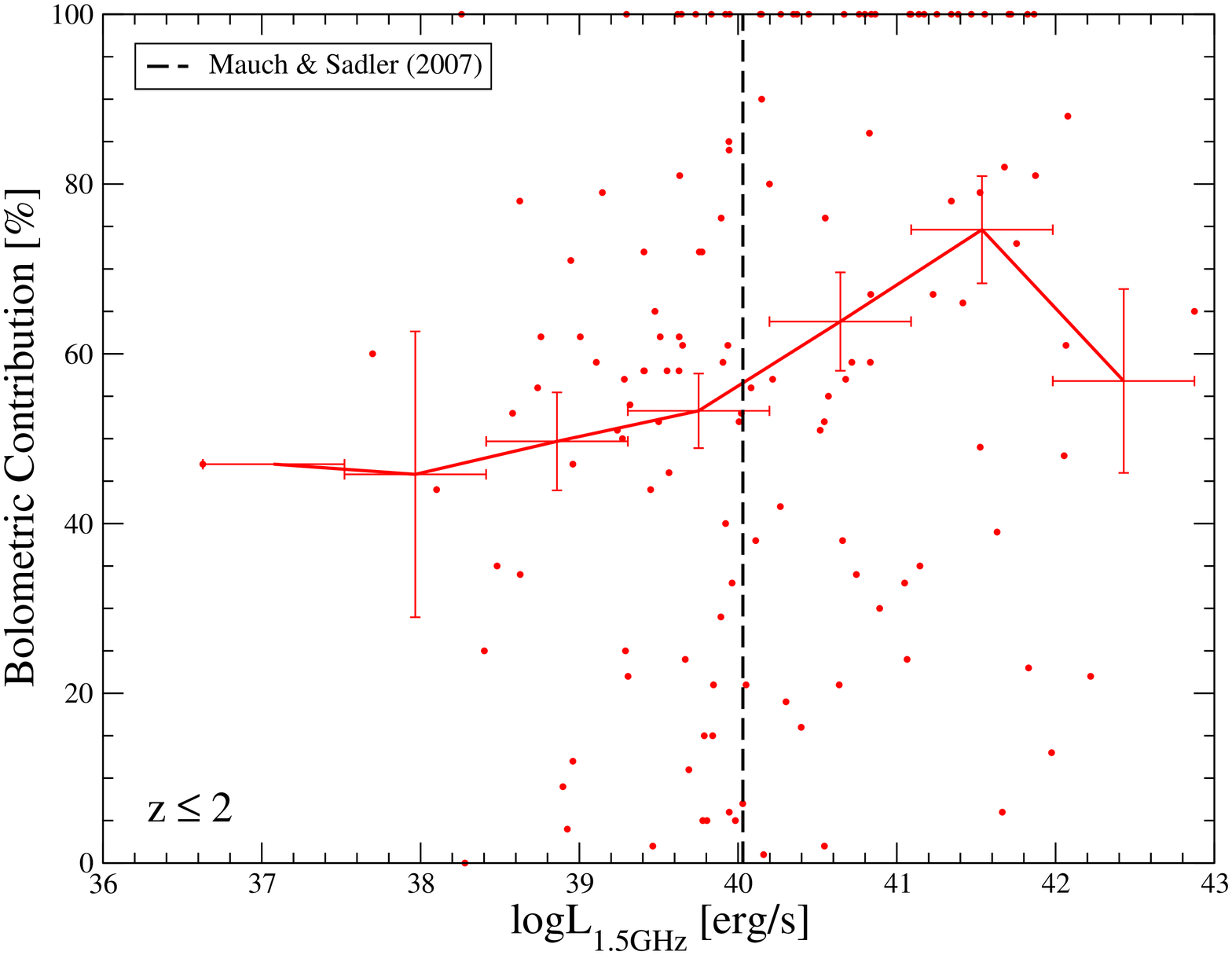}
\includegraphics[width=0.49\textwidth,angle=0]{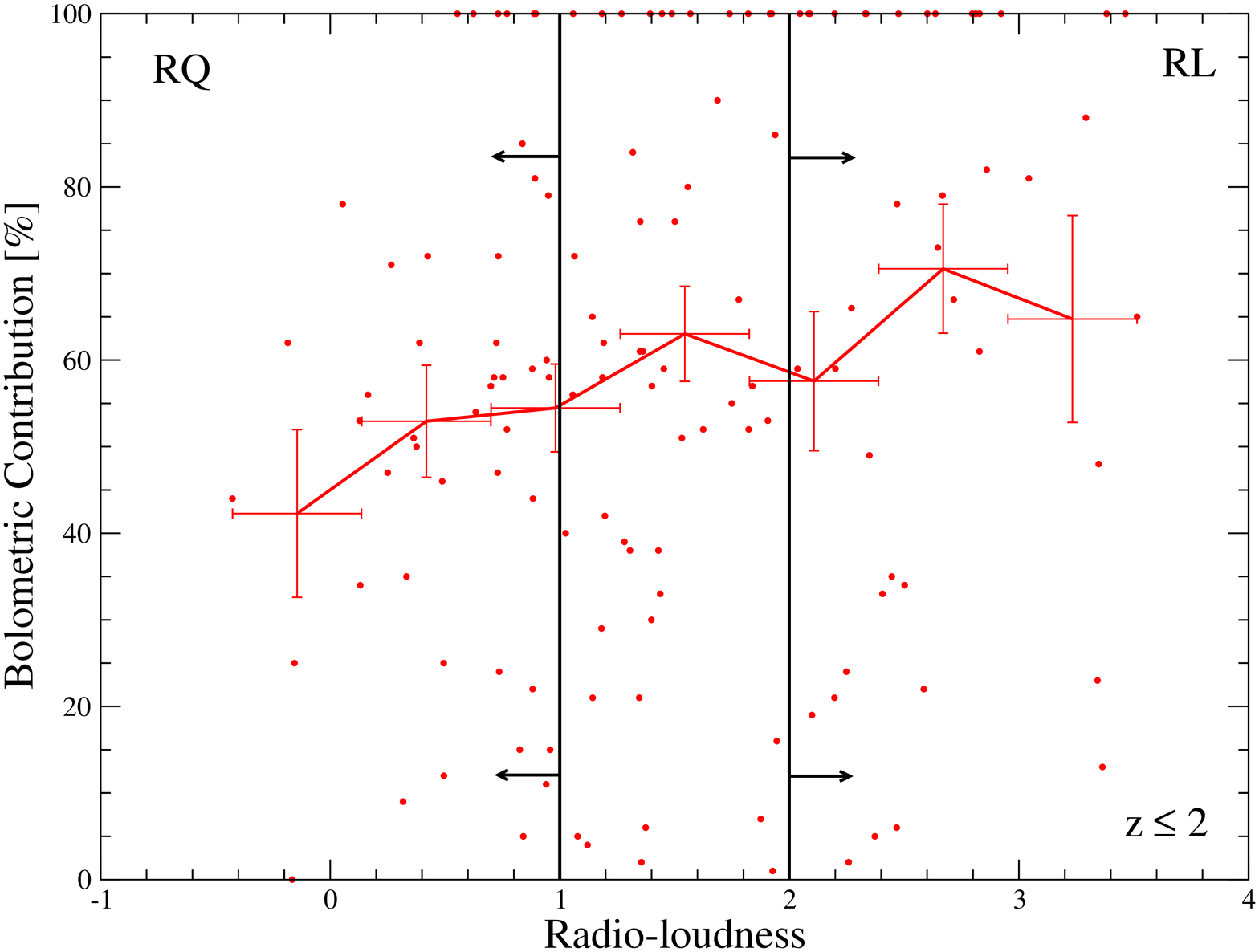}
\caption{AGN bolometric contribution as a function of 1.5 GHz luminosity $L_{1.5GHz}$ (left) and radio loudness $R_{i}$ (right). The red solid line shows average values of AGN fractional bolometric luminosity contribution over $L_{1.5GHz}$ bins. In the left panel with the vertical black dashed line is a transition luminosity taken from \citet{Mauch2007} (see text). In the right panel we also note the definitions of radio-loud and radio-quiet sources with vertical solid black lines. Horizontal error bars denote the respective bin width.}
\label{fig:AGN_SF_radio_Ri}
\end{center}
\end{figure}

In using the SED fitting results we can first of all see how the AGN and star-formation bolometric luminosity fractional contribution behaves as a function of both radio luminosity and radio-loudness. The results are shown in Fig. \ref{fig:AGN_SF_radio_Ri}.\\

Both for the highest radio luminosities and for the highest radio loudness values, sources have their bolometric luminosities dominated by an AGN component. These are the radio-AGN systems the star-formation properties of which we are interested in. The change between star-formation and AGN dominance\footnote{We note that if $\alpha$ is the AGN fractional bolometric contribution, then the star-formation fractional bolometric contribution is, by definition, $1-\alpha$.} in terms of bolometric luminosity agrees well with previous studies and lies roughly around $10^{40}$ erg/s or $10^{23}$ W/Hz (e.g., \citealt{Mauch2007}).\\
Several other points of interest arise from Fig. \ref{fig:AGN_SF_radio_Ri}. Even at the lowest radio luminosities, where star-formation should dominate, we find on average $\sim40\pm3\%$ contribution to the energy output from an AGN component. Similar behavior is seen for the most radio-quiet sources. Conversely, for the highest radio-luminosities $\sim30\pm3\%$ of the bolometric luminosity is contributed by star-formation. For the most radio-loud systems the star-formation contribution is further suppressed, with below $\sim20$\% fractional bolometric contribution. It should be noted that for the last bins of both radio luminosity and radio-loudness, we observed a change in the aforementioned trends. This is most probably an effect of low-number statistics (2 and 7 sources respectively contained in each of these two bins) combined with possible photometric redshift effects.\\ 

\begin{figure}[htbp]
\begin{center}
\includegraphics[width=0.49\textwidth,angle=0]{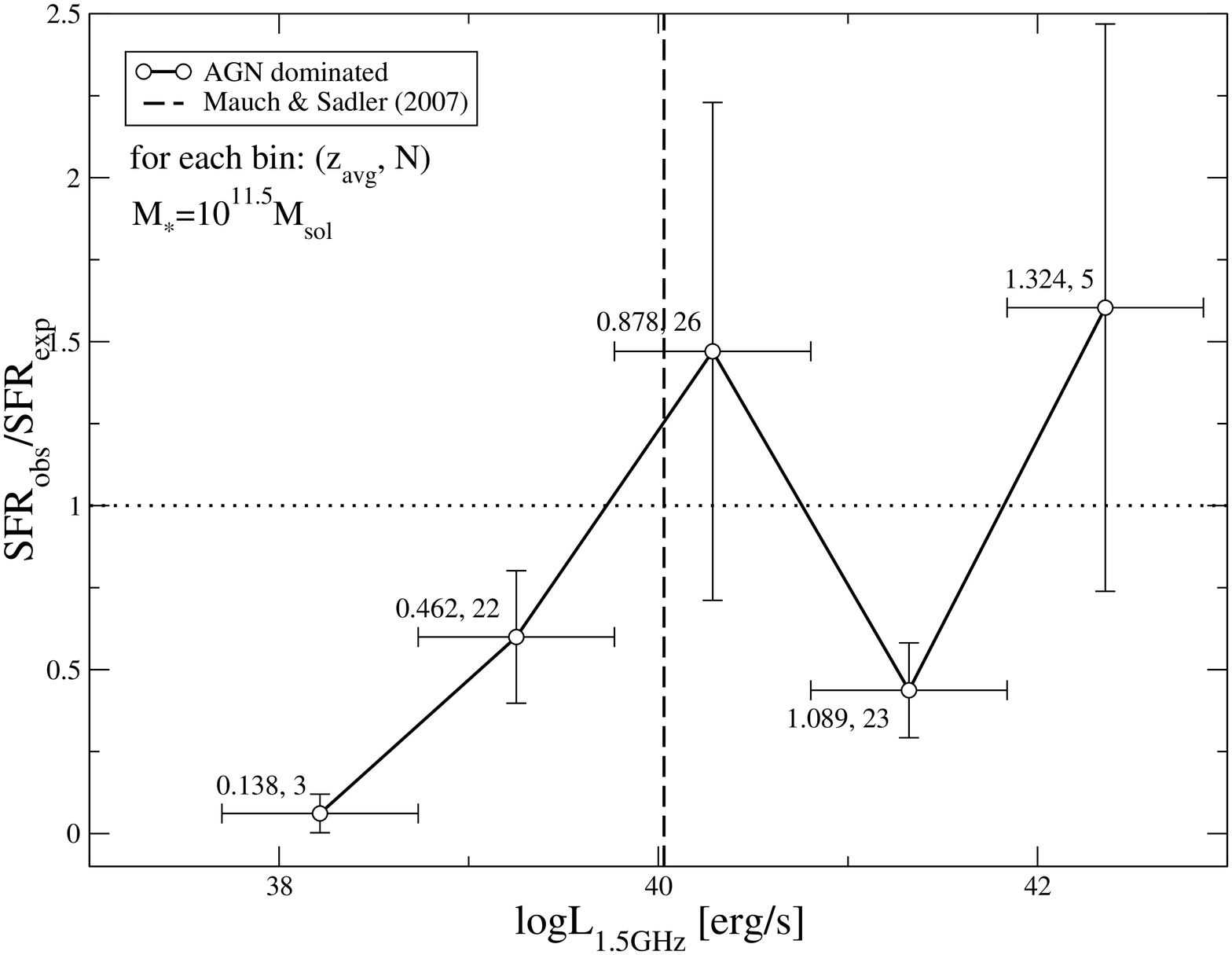}
\includegraphics[width=0.49\textwidth,angle=0]{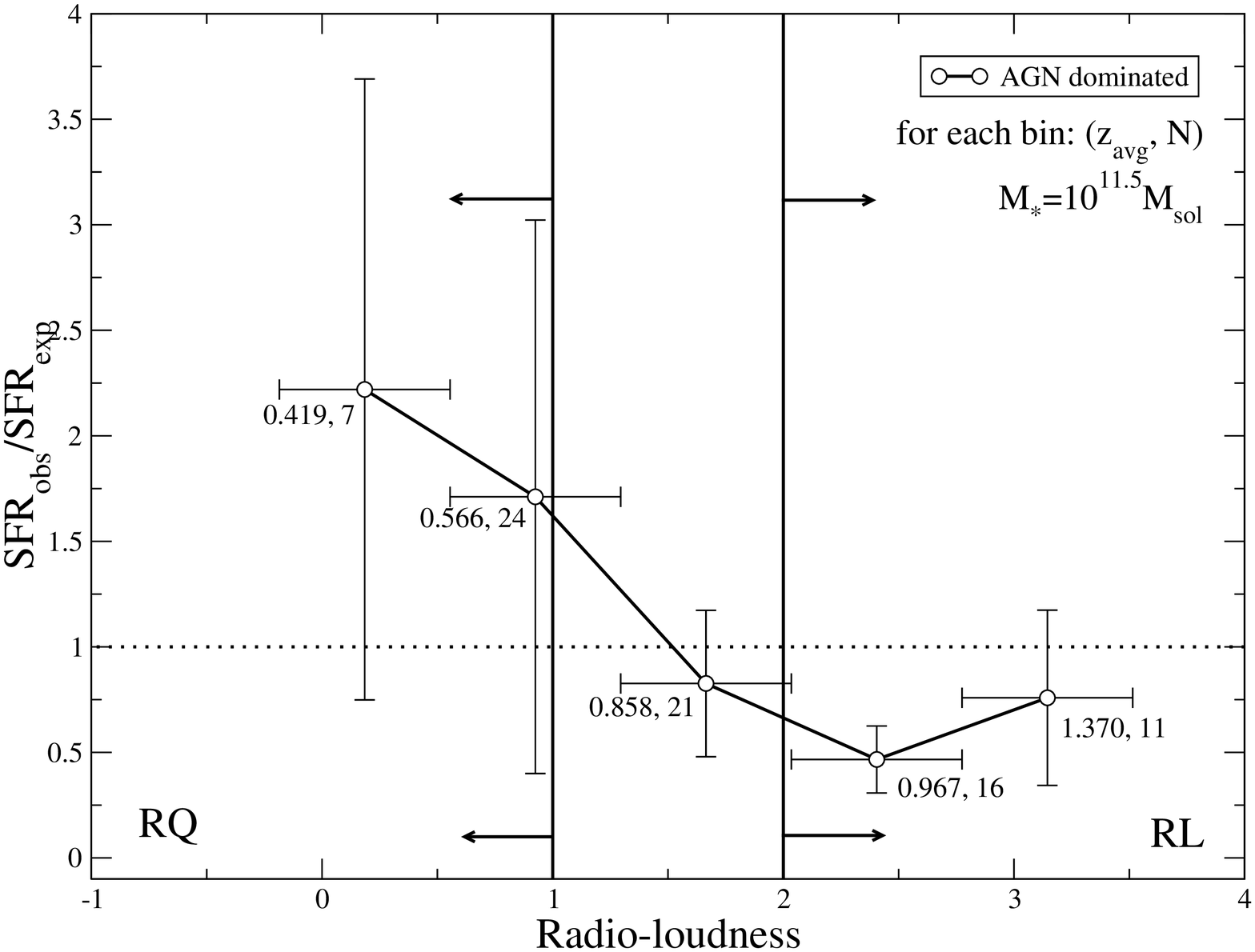}
\caption{Ratio between observed and expected star-formation rates (see text) as a function of $L_{1.5GHz}$ (left) and as a function of radio-loudness $R_{i}$ (right). We plot average values over luminosity and radio-loudness bins, respectively, for AGN dominated sources. Numbers denote the AGN dominated average redshifts and number of sources for each respective bin. The horizontal line separates the negative (below) and positive (above) feedback space. The vertical lines are the same as in Fig. 2.}
\label{fig:SFR_Lr_Ri_avg}
\end{center}
\end{figure}

\section {Star-formation quenching?}
\label{sec:feedback}

We use the star-formation component far-IR luminosity $L_{8-1000\mu m}$ from the SED fitting to calculate the SFR for the IR-radio sample using the relation from \citet{Kennicutt1998}. Although this is a theoretically derived relation, empirical calibrations of the relation between SFRs and $L_{8-1000\mu m}$ exist in the literature. These are usually within $30\%$ of the relation from \citet{Kennicutt1998} and therefore broadly agree with each other.\\
In Fig. \ref{fig:SFR_Lr_Ri_avg} the ratio between observed and expected SFRs per radio luminosity (left) and radio-loudness bins (right) is presented for the AGN dominated objects\footnote{We define AGN dominated objects such that their AGN fractional bolometric contribution is larger than 55\%.}. By using the empirical relation from \citet{Elbaz2011} that connects specific star-formation rate (sSFR) with z, we can calculate the expected SFRs for a passively evolving galaxy at the given redshift of each bin and assuming a stellar mass of $M_{\star}=10^{11.5}M_{sol}$ (mean stellar mass of radio-loud AGN, e.g., \citealt{Best2005}). We can thus calculate the ratios plotted in Fig. \ref{fig:SFR_Lr_Ri_avg}.\\
When looking at objects with luminosities around $\sim10^{41}$ erg/s, we see that with an average redshift of $\sim1$, they exhibit SFR below the expected levels. The same behavior is also seen when considering radio-loudness, with sources at $R_{i}\sim2.5$ exhibiting lower SFRs than expected given their average redshift. However, if we look at the last bin of both radio luminosity and radio-loudness, objects appear to show the expected amount of star-formation.\\ 

\section{Conclusions}
In conclusion, we see hints for lower SFRs than expected for radio-AGN dominated systems. However, if we consider the statistical errors together with the intrinsic scatter of the assumed sSFR to z empirical relation, we do not see a strong signature of negative feedback in radio-loud AGN. Conversely, our data clearly demonstrate the absence of positive feedback in radio-loud AGN. Estimation of individual stellar masses for each of our objects will allow us to differentiate between a negative or a non-existent feedback signature in these sources.\\

{\footnotesize
\noindent \emph{Acknowledgments:} MK acknowledges the support from the Creative Research Initiative program, No. 2010-0000712, of the National Research Foundation of Korea (NRFK) funded by the Korea government(MEST). 

%\begin{thebibliography}{99}
%
%  \bibitem{...} ....
%
%\end{thebibliography}
\bibsep-0.7ex
\bibliographystyle{aa}
\bibliography{bibtex}

\begin{thebibliography}{29}
\expandafter\ifx\csname natexlab\endcsname\relax\def\natexlab#1{#1}\fi

\bibitem[{{Arnouts} {et~al.}(1999){Arnouts}, {Cristiani}, {Moscardini},
  {Matarrese}, {Lucchin}, {Fontana}, \& {Giallongo}}]{Arnouts1999}
{Arnouts}, S., {Cristiani}, S., {Moscardini}, L., {et~al.} 1999, \mnras, 310,
  540

\bibitem[{{Best} \& {Heckman}(2012)}]{Best2012}
{Best}, P.~N. \& {Heckman}, T.~M. 2012, \mnras, 421, 1569

\bibitem[{{Best} {et~al.}(2005){Best}, {Kauffmann}, {Heckman}, {Brinchmann},
  {Charlot}, {Ivezi{\'c}}, \& {White}}]{Best2005}
{Best}, P.~N., {Kauffmann}, G., {Heckman}, T.~M., {et~al.} 2005, \mnras, 362,
  25

\bibitem[{{Best} {et~al.}(2007){Best}, {von der Linden}, {Kauffmann},
  {Heckman}, \& {Kaiser}}]{Best2007}
{Best}, P.~N., {von der Linden}, A., {Kauffmann}, G., {Heckman}, T.~M., \&
  {Kaiser}, C.~R. 2007, \mnras, 379, 894

\bibitem[{{Bicknell} {et~al.}(2000){Bicknell}, {Sutherland}, {van Breugel},
  {Dopita}, {Dey}, \& {Miley}}]{Bicknell2000}
{Bicknell}, G.~V., {Sutherland}, R.~S., {van Breugel}, W.~J.~M., {et~al.} 2000,
  \apj, 540, 678

\bibitem[{{Condon}(1992)}]{Condon1992}
{Condon}, J.~J. 1992, \araa, 30, 575

\bibitem[{{Croton} {et~al.}(2006){Croton}, {Springel}, {White}, {De Lucia},
  {Frenk}, {Gao}, {Jenkins}, {Kauffmann}, {Navarro}, \& {Yoshida}}]{Croton2006}
{Croton}, D.~J., {Springel}, V., {White}, S.~D.~M., {et~al.} 2006, \mnras, 365,
  11

\bibitem[{{Downes} {et~al.}(1986){Downes}, {Peacock}, {Savage}, \&
  {Carrie}}]{Downes1986}
{Downes}, A.~J.~B., {Peacock}, J.~A., {Savage}, A., \& {Carrie}, D.~R. 1986,
  \mnras, 218, 31

\bibitem[{{Elbaz} {et~al.}(2011){Elbaz}, {Dickinson}, {Hwang},
  {D{\'{\i}}az-Santos}, {Magdis}, {Magnelli}, {Le Borgne}, {Galliano},
  {Pannella}, {Chanial}, {Armus}, {Charmandaris}, {Daddi}, {Aussel}, {Popesso},
  {Kartaltepe}, {Altieri}, {Valtchanov}, {Coia}, {Dannerbauer}, {Dasyra},
  {Leiton}, {Mazzarella}, {Alexander}, {Buat}, {Burgarella}, {Chary}, {Gilli},
  {Ivison}, {Juneau}, {Le Floc'h}, {Lutz}, {Morrison}, {Mullaney}, {Murphy},
  {Pope}, {Scott}, {Brodwin}, {Calzetti}, {Cesarsky}, {Charlot}, {Dole},
  {Eisenhardt}, {Ferguson}, {F{\"o}rster Schreiber}, {Frayer}, {Giavalisco},
  {Huynh}, {Koekemoer}, {Papovich}, {Reddy}, {Surace}, {Teplitz}, {Yun}, \&
  {Wilson}}]{Elbaz2011}
{Elbaz}, D., {Dickinson}, M., {Hwang}, H.~S., {et~al.} 2011, \aap, 533, A119

\bibitem[{{Ferrarese} \& {Merritt}(2000)}]{Ferrarese2000}
{Ferrarese}, L. \& {Merritt}, D. 2000, \apjl, 539, L9

\bibitem[{{Hwang} {et~al.}(2007){Hwang}, {Lee}, {Lee}, {Im}, {Kim},
  {Matsuhara}, {Wada}, {Oyabu}, {Pak}, {Chun}, {Watarai}, {Nakagawa},
  {Pearson}, {Takagi}, {Hanami}, \& {White}}]{Hwang2007}
{Hwang}, N., {Lee}, M.~G., {Lee}, H.~M., {et~al.} 2007, \apjs, 172, 583

\bibitem[{{Ilbert} {et~al.}(2006){Ilbert}, {Arnouts}, {McCracken},
  {Bolzonella}, {Bertin}, {Le F{\`e}vre}, {Mellier}, {Zamorani}, {Pell{\`o}},
  {Iovino}, {Tresse}, {Le Brun}, {Bottini}, {Garilli}, {Maccagni}, {Picat},
  {Scaramella}, {Scodeggio}, {Vettolani}, {Zanichelli}, {Adami}, {Bardelli},
  {Cappi}, {Charlot}, {Ciliegi}, {Contini}, {Cucciati}, {Foucaud}, {Franzetti},
  {Gavignaud}, {Guzzo}, {Marano}, {Marinoni}, {Mazure}, {Meneux}, {Merighi},
  {Paltani}, {Pollo}, {Pozzetti}, {Radovich}, {Zucca}, {Bondi}, {Bongiorno},
  {Busarello}, {de La Torre}, {Gregorini}, {Lamareille}, {Mathez}, {Merluzzi},
  {Ripepi}, {Rizzo}, \& {Vergani}}]{Ilbert2006}
{Ilbert}, O., {Arnouts}, S., {McCracken}, H.~J., {et~al.} 2006, \aap, 457, 841

\bibitem[{{Ivezi{\'c}} {et~al.}(2002){Ivezi{\'c}}, {Menou}, {Knapp}, {Strauss},
  {Lupton}, {Vanden Berk}, {Richards}, {Tremonti}, {Weinstein}, {Anderson},
  {Bahcall}, {Becker}, {Bernardi}, {Blanton}, {Eisenstein}, {Fan},
  {Finkbeiner}, {Finlator}, {Frieman}, {Gunn}, {Hall}, {Kim}, {Kinkhabwala},
  {Narayanan}, {Rockosi}, {Schlegel}, {Schneider}, {Strateva}, {SubbaRao},
  {Thakar}, {Voges}, {White}, {Yanny}, {Brinkmann}, {Doi}, {Fukugita},
  {Hennessy}, {Munn}, {Nichol}, \& {York}}]{Ivezic2002}
{Ivezi{\'c}}, {\v Z}., {Menou}, K., {Knapp}, G.~R., {et~al.} 2002, \aj, 124,
  2364

\bibitem[{{Jeon} {et~al.}(2010){Jeon}, {Im}, {Ibrahimov}, {Lee}, {Lee}, \&
  {Lee}}]{Jeon2010}
{Jeon}, Y., {Im}, M., {Ibrahimov}, M., {et~al.} 2010, \apjs, 190, 166

\bibitem[{{Kalfountzou} {et~al.}(2012){Kalfountzou}, {Jarvis}, {Bonfield}, \&
  {Hardcastle}}]{Kalfountzou2012}
{Kalfountzou}, E., {Jarvis}, M.~J., {Bonfield}, D.~G., \& {Hardcastle}, M.~J.
  2012, ArXiv e-prints

\bibitem[{{Kennicutt}(1998)}]{Kennicutt1998}
{Kennicutt}, Jr., R.~C. 1998, \araa, 36, 189

\bibitem[{{Kim} {et~al.}(2012){Kim}, {Lee}, {Matsuhara}, {Wada}, {Oyabu}, {Im},
  {Jeon}, {Kang}, {Ko}, {Lee}, {Takagi}, {Pearson}, {White}, {Jeong},
  {Serjeant}, {Nakagawa}, {Ohyama}, {Goto}, {Takeuchi}, {Pollo}, {Solarz}, \&
  {Pepiak}}]{Kim2012}
{Kim}, S.~J., {Lee}, H.~M., {Matsuhara}, H., {et~al.} 2012, ArXiv e-prints

\bibitem[{{Kormendy} \& {Richstone}(1995)}]{Kormendy1995}
{Kormendy}, J. \& {Richstone}, D. 1995, \araa, 33, 581

\bibitem[{{Magorrian} {et~al.}(1998){Magorrian}, {Tremaine}, {Richstone},
  {Bender}, {Bower}, {Dressler}, {Faber}, {Gebhardt}, {Green}, {Grillmair},
  {Kormendy}, \& {Lauer}}]{Magorrian1998}
{Magorrian}, J., {Tremaine}, S., {Richstone}, D., {et~al.} 1998, \aj, 115, 2285

\bibitem[{{Mauch} \& {Sadler}(2007)}]{Mauch2007}
{Mauch}, T. \& {Sadler}, E.~M. 2007, \mnras, 375, 931

\bibitem[{{McNamara} {et~al.}(2005){McNamara}, {Nulsen}, {Wise}, {Rafferty},
  {Carilli}, {Sarazin}, \& {Blanton}}]{McNamara2005}
{McNamara}, B.~R., {Nulsen}, P.~E.~J., {Wise}, M.~W., {et~al.} 2005, \nat, 433,
  45

\bibitem[{{Murakami} {et~al.}(2007){Murakami}, {Baba}, {Barthel}, {Clements},
  {Cohen}, {Doi}, {Enya}, {Figueredo}, {Fujishiro}, {Fujiwara}, {Fujiwara},
  {Garcia-Lario}, {Goto}, {Hasegawa}, {Hibi}, {Hirao}, {Hiromoto}, {Hong},
  {Imai}, {Ishigaki}, {Ishiguro}, {Ishihara}, {Ita}, {Jeong}, {Jeong},
  {Kaneda}, {Kataza}, {Kawada}, {Kawai}, {Kawamura}, {Kessler}, {Kester},
  {Kii}, {Kim}, {Kim}, {Kobayashi}, {Koo}, {Kwon}, {Lee}, {Lorente}, {Makiuti},
  {Matsuhara}, {Matsumoto}, {Matsuo}, {Matsuura}, {M{\"u}ller}, {Murakami},
  {Nagata}, {Nakagawa}, {Naoi}, {Narita}, {Noda}, {Oh}, {Ohnishi}, {Ohyama},
  {Okada}, {Okuda}, {Oliver}, {Onaka}, {Ootsubo}, {Oyabu}, {Pak}, {Park},
  {Pearson}, {Rowan-Robinson}, {Saito}, {Sakon}, {Salama}, {Sato}, {Savage},
  {Serjeant}, {Shibai}, {Shirahata}, {Sohn}, {Suzuki}, {Takagi}, {Takahashi},
  {Tanab{\'e}}, {Takeuchi}, {Takita}, {Thomson}, {Uemizu}, {Ueno}, {Usui},
  {Verdugo}, {Wada}, {Wang}, {Watabe}, {Watarai}, {White}, {Yamamura},
  {Yamauchi}, \& {Yasuda}}]{Murakami2007}
{Murakami}, H., {Baba}, H., {Barthel}, P., {et~al.} 2007, \pasj, 59, 369

\bibitem[{{Onaka} {et~al.}(2007){Onaka}, {Matsuhara}, {Wada}, {Fujishiro},
  {Fujiwara}, {Ishigaki}, {Ishihara}, {Ita}, {Kataza}, {Kim}, {Matsumoto},
  {Murakami}, {Ohyama}, {Oyabu}, {Sakon}, {Tanab{\'e}}, {Takagi}, {Uemizu},
  {Ueno}, {Usui}, {Watarai}, {Cohen}, {Enya}, {Ootsubo}, {Pearson}, {Takeyama},
  {Yamamuro}, \& {Ikeda}}]{Onaka2007}
{Onaka}, T., {Matsuhara}, H., {Wada}, T., {et~al.} 2007, \pasj, 59, 401

\bibitem[{{Polletta} {et~al.}(2007){Polletta}, {Tajer}, {Maraschi},
  {Trinchieri}, {Lonsdale}, {Chiappetti}, {Andreon}, {Pierre}, {Le F{\`e}vre},
  {Zamorani}, {Maccagni}, {Garcet}, {Surdej}, {Franceschini}, {Alloin},
  {Shupe}, {Surace}, {Fang}, {Rowan-Robinson}, {Smith}, \&
  {Tresse}}]{Polletta2007}
{Polletta}, M., {Tajer}, M., {Maraschi}, L., {et~al.} 2007, \apj, 663, 81

\bibitem[{{Schlegel} {et~al.}(1998){Schlegel}, {Finkbeiner}, \&
  {Davis}}]{Schlegel1998}
{Schlegel}, D.~J., {Finkbeiner}, D.~P., \& {Davis}, M. 1998, \apj, 500, 525

\bibitem[{{Spergel} {et~al.}(2003){Spergel}, {Verde}, {Peiris}, {Komatsu},
  {Nolta}, {Bennett}, {Halpern}, {Hinshaw}, {Jarosik}, {Kogut}, {Limon},
  {Meyer}, {Page}, {Tucker}, {Weiland}, {Wollack}, \& {Wright}}]{Spergel2003}
{Spergel}, D.~N., {Verde}, L., {Peiris}, H.~V., {et~al.} 2003, \apjs, 148, 175

\bibitem[{{Trichas} {et~al.}(2012){Trichas}, {Green}, {Silverman}, {Aldcroft},
  {Barkhouse}, {Cameron}, {Constantin}, {Ellison}, {Foltz}, {Haggard},
  {Jannuzi}, {Kim}, {Marshall}, {Mossman}, {P{\'e}rez}, {Romero-Colmenero},
  {Ruiz}, {Smith}, {Smith}, {Torres}, {Wik}, {Wilkes}, \&
  {Wolfgang}}]{Trichas2012}
{Trichas}, M., {Green}, P.~J., {Silverman}, J.~D., {et~al.} 2012, \apjs, 200,
  17

\bibitem[{{White} {et~al.}(2010){White}, {Pearson}, {Braun}, {Serjeant},
  {Matsuhara}, {Takagi}, {Nakagawa}, {Shipman}, {Barthel}, {Hwang}, {Lee},
  {Lee}, {Im}, {Wada}, {Oyabu}, {Pak}, {Chun}, {Hanami}, {Goto}, \&
  {Oliver}}]{White2010}
{White}, G.~J., {Pearson}, C., {Braun}, R., {et~al.} 2010, \aap, 517, A54

\bibitem[{{Wright} {et~al.}(2010){Wright}, {Eisenhardt}, {Mainzer}, {Ressler},
  {Cutri}, {Jarrett}, {Kirkpatrick}, {Padgett}, {McMillan}, {Skrutskie},
  {Stanford}, {Cohen}, {Walker}, {Mather}, {Leisawitz}, {Gautier}, {McLean},
  {Benford}, {Lonsdale}, {Blain}, {Mendez}, {Irace}, {Duval}, {Liu}, {Royer},
  {Heinrichsen}, {Howard}, {Shannon}, {Kendall}, {Walsh}, {Larsen}, {Cardon},
  {Schick}, {Schwalm}, {Abid}, {Fabinsky}, {Naes}, \& {Tsai}}]{Wright2010}
{Wright}, E.~L., {Eisenhardt}, P.~R.~M., {Mainzer}, A.~K., {et~al.} 2010, \aj,
  140, 1868

\end{thebibliography}
}

\end{document}